\documentclass[twocolumn]{aastex62}

\shorttitle{Coherent ICS in FRBs}
\shortauthors{Zhang}

\begin{document}


\title{\bf  Coherent inverse Compton scattering by bunches in fast radio bursts}


\author{Bing Zhang }
\affil{Department of Physics and Astronomy, University of Nevada, Las Vegas, Las Vegas, NV 89154,
bing.zhang@unlv.edu}

\begin{abstract}
The extremely high brightness temperature of fast radio bursts (FRBs) requires that their emission mechanism must be ``coherent'', either through concerted particle emission by bunches or through an exponential growth of a plasma wave mode or radiation amplitude via certain maser mechanisms. The bunching mechanism has been mostly discussed within the context of curvature radiation or cyclotron/synchrotron radiation. Here we propose a family of model invoking coherent inverse Compton scattering (ICS) of bunched particles that may operate within or just outside of the magnetosphere of a flaring magnetar. Crustal oscillations during the flaring event may excite low-frequency electromagnetic waves near the magnetar surface. The X-mode of these waves could penetrate through the magnetosphere. Bunched relativistic particles in the charge starved region inside the magnetosphere or in the current sheet outside of the magnetosphere would upscatter these low-frequency waves to produce GHz emission to power  FRBs. The ICS mechanism has a much larger emission power for individual electrons than curvature radiation. This greatly reduces the required degree of coherence in bunches, alleviating several criticisms to the bunching mechanism raised in the context of curvature radiation. The emission is $\sim 100\%$ linearly polarized (with the possibility of developing circular polarization) with a constant or varying polarization angle across each burst. The mechanism can account for a narrow-band spectrum and a frequency downdrifting pattern, as commonly observed in repeating FRBs. 
\end{abstract}


\section{Introduction} \label{sec:intro}

Fast radio bursts (FRBs) \citep{lorimer07,petroff19,cordes19,zhang20b} have a brightness temperature
\begin{eqnarray}
        T_b & \simeq & \frac{{\cal S}_{\nu,p} D_{\rm A}^2}{2 \pi k_{\rm B} (\nu \Delta t)^2} = (1.2\times 10^{36}  \ {\rm K} )\nonumber \\
        & \times & D_{\rm A,28}^{2} ({\cal S}_{\nu,p}/{\rm Jy}) \nu_9^{-2} \Delta t_{-3}^{-2},
        \label{eq:Tb}
\end{eqnarray}
where $k_{\rm B}$ is Boltzmann constant, $S_{\nu,p}$ is specific flux at the peak time, $\nu$ is observing frequency, $\Delta t$ is variability timescale, $D_{\rm A}$ is angular distance of the source, and the convention $Q_n = Q/10^n$ has been adopted in cgs units throughout the paper. This is much greater than the maximum brightness temperature for an incoherent emitting source 
\begin{equation}
  T_{\rm b,max}^{\rm incoh} \simeq \Gamma \gamma m_e c^2 / k_{\rm B} = (5.9\times 10^{13} \ {\rm K}) \Gamma_2 \gamma_{2},
\end{equation}
where $m_e$, $c$, $k$ are the fundamental constants electron mass, speed of light and Boltzman constant, respectively, $\Gamma$ is the bulk Lorentz factor of the emitter towards earth (taken as unity if the source is not moving with a relativistic speed), and $\gamma = {\rm max} (\gamma_m, \gamma_a)$ is the characteristic electron Lorentz factor in the emission region, which, for a synchrotron source, is the greater of the minimum injection Lorentz factor $\gamma_m$ and the Lorentz factor $\gamma_a$ corresponding to synchrotron self-absorption \citep{kumar15}.  This suggests that FRB emission mechanism must be coherent. 

Many coherent radiation mechanisms have been explored to interpret the emission of radio pulsars, which also have the brightness temperature $T_b \gg  T_{\rm b,max}^{\rm incoh}$ (but is about 10 orders of magnitude lower than that of FRBs). In general, these mechanisms can be grouped into three types \citep[e.g.][]{melrose78}: coherent radiation by bunches, intrinsic growth of plasma wave modes, and maser mechanism. The second and third mechanisms were also termed as ``plasma maser'' and ``vacuum maser'' mechanisms, respectively. Some of these mechanisms have been reinvented to interpret FRB coherent emission \citep[e.g.][for surveys of various mechanisms discussed in the literature]{lu18,zhang20b,xiao21,lyubarsky21}.

This paper mainly concerns about the first type of coherent mechanism, namely, coherent radiation by bunches, also called the ``antenna'' mechanism. Within this mechanism, charged particles are clustered in both position and momentum spaces and emit as a macroscopic charge. The emission power of the bunch is $P_b \simeq N_{e,b}^2 P_e$, where $P_e$ is the power of individual electrons, and $N_{e,b}$ is the total number of net charges in the bunch. Within the FRB context, a widely discussed possibility is coherent curvature radiation by bunches \citep{katz14,kumar17,yangzhang18,lu18,wang19,kumar20a,lu20,yang20c,wanglai20,cooper21}. Another related mechanism is the so-called ``synchrotron maser'' mechanism in 90$^{\rm o}$ magnetized relativistic shocks, which in fact invoke cyclotron-radiating bunches of charged particles in momentum space \citep{lyubarsky14,beloborodov17,beloborodov20,metzger19,plotnikov19,margalit20}. These mechanisms can interpret some FRB emission properties but also suffer from criticisms in theoretical and/or observational aspects \citep[e.g.][]{zhang20b,lyubarsky21}.

In this paper, we discuss another family of bunching coherent mechanisms, namely, coherent inverse Compton scattering (ICS) by bunches. Before getting into the details of the model, it is informative to clarify the meaning of ``bunches'' discussed in this paper. In the pulsar literature, ``bunches'' are usually treated as giant charges surrounded by a background plasma \citep[e.g.][]{gil04}. The mechanism was criticized for the the formation and maintenance mechanisms of such bunches \citep[e.g.][]{melrose78}. Also the surrounding plasma works against the coherence of the bunch and suppresses the coherent emission \citep{gil04,lyubarsky21}. For the curvature radiation mechanism widely discussed in the literature, since the radiation power of individual particles is low, highly coherent bunches are required and these criticisms may be relevant \citep[but see][]{melikidze00,qu21}.  However, as will be shown below, in the case of ICS emission, since the emission power of individual particles is much higher, the required coherence for bunches is much reduced. Rather than distinct giant charges surrounded by an ambient medium, the bunches discussed in this paper are merely charge density fluctuations in an relativistic particle outflow (see also \citealt{yangzhang18}).

\begin{figure}
\plotone{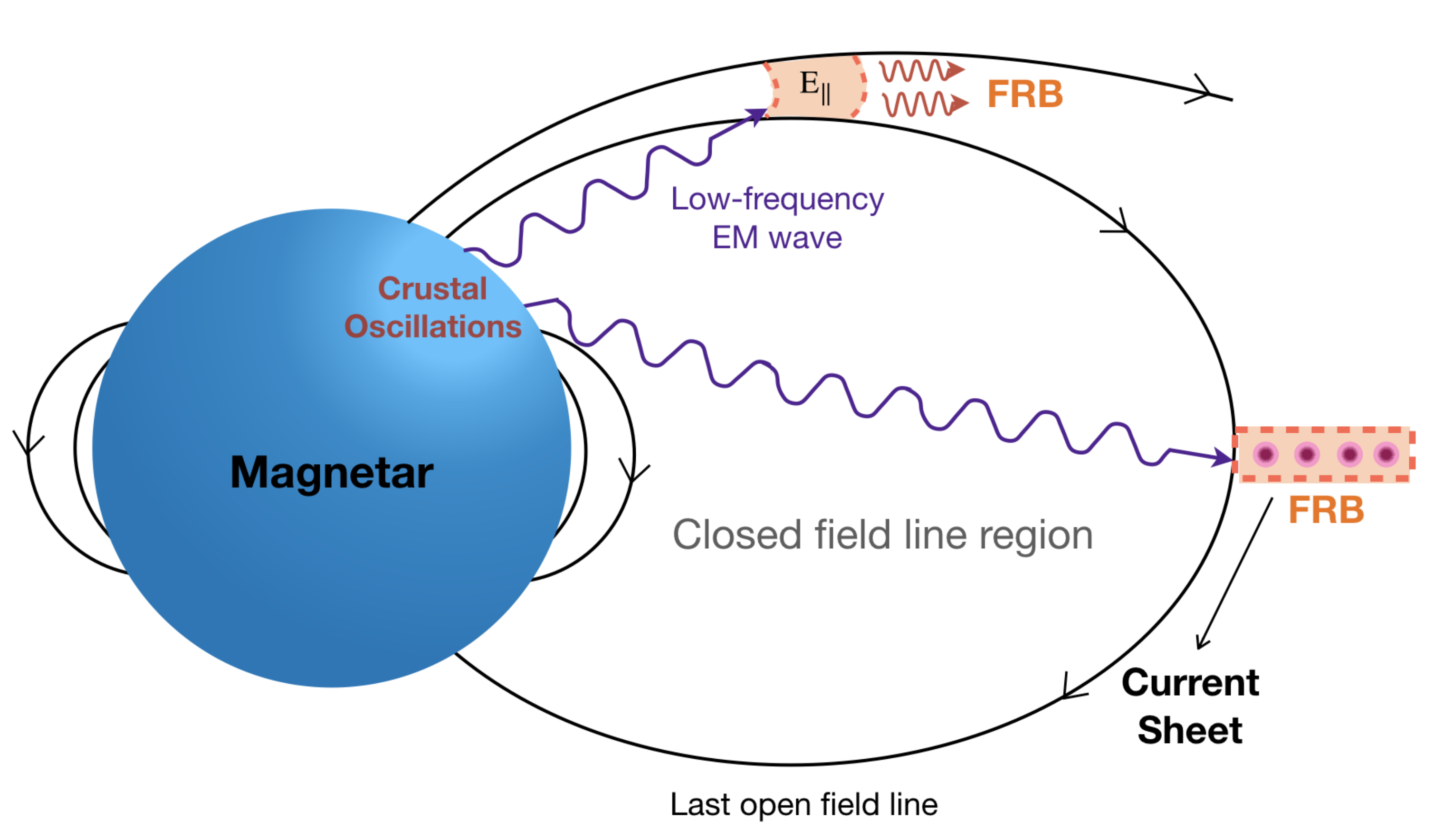}
\caption{The geometric sketch of the proposed model. Charges in the magnetosphere near the crustal oscillation region send out low-frequency electromagnetic waves which propagate across the closed field line region to reach the emission sites, either in the outer magnetosphere in the open field line region where an $E_\parallel$ is developed, or in the current sheet region outside the magnetosphere. Relativistic particles in the emission region upscatter the low-frequency waves and produce FRBs. The FRB emission direction is along the local magnetic field lines in the first scenario and is out of page in the second scenario. }
\label{fig:cartoon}
\end{figure}

\section{Coherent ICS by bunches}

\subsection{General picture}
The general picture of the hereby discussed mechanism is the following (Figure \ref{fig:cartoon}): Sudden cracking of the curst of a magnetar excites crustal quakes and plasma oscillations near the surface of the magnetar.  Low-frequency electromagnetic waves with angular frequency $\omega_0$ would be generated by the coherently oscillating charged particles and would propagate across the magnetosphere. The waves encounter bunches of relativistic particles with a typical Lorentz factor $\gamma$ and a net charge $N_{e,b} e$, which moves along magnetic field lines in the outer  magnetosphere or in current sheets outside of the magnetosphere. In the rest-frame of the particle bunch, the incident wave frequency is boosted to $\sim \gamma (1-\beta\cos\theta_i) \nu_0$. The bunched particles are induced to oscillate at the same frequency in the comoving frame, which is transferred to a lab-frame outgoing ICS frequency 
\begin{equation}
 \nu \simeq \gamma^2 \nu_0 (1-\beta\cos\theta_i)= (1 \ {\rm GHz}) \ \gamma_{2.5}^2 \nu_{0,4} (1-\beta\cos\theta_i),
 \label{eq:nu}
\end{equation}
where $\nu_0 = \omega_0 / 2\pi $ and $\theta_i$ is the angle between the incident photon momentum and the electron momentum. Similar to curvature radiation, such an ICS mechanism does not depend intrinsically on the dispersive properties of the plasma and may be treated assuming vacuum wave properties. Different from the curvature radiation mechanism which appeals to the curved trajectory of charged particles for acceleration, the ICS mechanism invokes an oscillating electromagnetic field from the low frequency waves to accelerate bunched particles.  

Such a mechanism is not new in the pulsar literature. Linear acceleration emission (LAE) invokes an oscillating $E_\parallel$ along the direction of particle motion \citep[e.g.][]{melrose78,rowe95}. A free-electron laser (FEL) invokes interaction between an electromagnetic disturbance (wiggler) with a relativistically moving bunch \citep[e.g.][]{fung04} to power pulsar radio emission or FRB emission \cite{lyutikov21}. \cite{lyubarsky96} introduced induced upscattering of longitudinal oscillations to interpret pulsar radio emission. \cite{qiao98} considered low-frequency electromagnetic waves generated from regular discharges of an inner vacuum  gap in the pulsar polar cap region and showed that ICS off these waves may reproduce many observational features of pulsar radio emission \citep{qiao01,xu00}.

\subsection{Low frequency waves: generation and propagation}
 
Since FRBs are rarely produced from their sources (e.g. the Galactic magnetar SGR 1935+2154 only produced one detected FRB since its discovery, \citealt{CHIME-SGR,STARE2-SGR}, while many X-ray bursts have been produced without association with any FRBs, \citealt{lin20}), it is reasonable to assume that something extraordinary must have happened to trigger an FRB. We envisage the low-frequency electromagnetic waves generated near the surface of the neutron star as one condition to produce FRBs. This may be related to violent oscillations of charged particles near the surface triggered by, e.g. strong oscillations of the neutron star crust during a starquake \citep[e.g.][]{thompson01,beloborodov07,wang18,dehman20,yangzhang21}. The bulk of the global oscillation energy would be carried by Alfv\'en waves that propagate along the magnetic field lines in the form a magnetic pulse, which would deposit energy in the form of particle energy and radiation at a large distance, as envisaged in most of the FRB models \citep[e.g.][]{kumar20a,lyubarsky20,yuan20}. However, the same shearing oscillations of the crust would induce oscillations of charges in the near-surface magnetosphere in the direction perpendicular to magnetic field lines, so that a small fraction of the oscillation energy may be radiated by these charges coherently and nearly isotropically as X-mode electromagnetic waves with angular frequency $\omega_0$, similar to  an antenna in a radio station on Earth. Strictly speaking, the electromagnetic waves discussed here are electromagnetic modes propagating in a plasma, which may be equivalent to some forms of fast magnetosonic waves. However, since the magnetization factor $\sigma$ is extremely high in the inner magnetosphere, the magnetosonic waves all propagate at essentially the speed of light and would behave like electromagnetic waves in vacuum with a proper dispersion relation. Since the electromagnetic waves do not carry the bulk of FRB energy but rather provide seed photons for particles to upscatter at a large radius,  the brightness temperature of the low-frequency coherent emission does not need to be very high. In this model, the low frequency waves should at least last several milliseconds (the duration of the FRB), but could last longer (in which case the FRB duration would be defined by the duration of the magnetic pulse dissipation leading to coherent ICS emission by relativistic particles). 

For a cold, non-relativistic plasma, the X-mode electromagnetic waves can propagate in the frequency regimes $\omega > \omega_{\rm R}$, $\omega_{\rm L} < \omega < \omega_{\rm uh}$, and $\omega < \omega_{\rm lh}$, where $\omega_{\rm R}$, $\omega_{\rm L}$ are the cutoff frequencies for the R-mode and L-mode and $\omega_{\rm uh}$ and $\omega_{\rm lh}$ are the upper and lower hybrid principle resonant frequencies, respectively \citep{boyd03}. For an electron-positron pair plasma relevant for a pulsar/magnetar magnetosphere, one has $\omega_{\rm R} = \omega_{\rm uh} = \omega_{\rm L} = \sqrt{\omega_p^2 + \omega_{\rm B}^2}$ and $\omega_{\rm lh} = \omega_{\rm B}$, so that the X-mode electromagnetic waves are free to propagate at $\omega > \sqrt{\omega_p^2 + \omega_{\rm B}^2}$ or $\omega < \omega_{\rm B}$ (see Appendix). For a magnetar with surface magnetic field $B_s = 10^{15} {\rm G} B_{s,15}$ and rotation period $P$, the Goldreich-Julian charge number density \citep{goldreich69,ruderman75} is 
\begin{equation}
n_{\rm GJ} \simeq \frac{\Omega B}{2\pi c e} \simeq (6.9\times 10^7 \ {\rm cm^{-3}}) B_{s,15} P^{-1}\hat  r_2^{-3},
\label{eq:ngj}
\end{equation}
where $\Omega=2\pi/P$ is the angular frequency, and
\begin{equation}
\hat r \equiv \frac{r}{R}
\end{equation} 
is the radius $r$ normalized to the neutron star radius $R=10^6 {\rm cm} R_6$. The plasma number density can be estimated as
\begin{equation}
n = \xi n_{\rm GJ},
\label{eq:n}
\end{equation}
where $\xi$ is the multiplicity parameter due to electron-positron pair production. As a result, the plasma frequency and the Larmor frequency can be estimated as
\begin{eqnarray}
 \omega_p & = & \left(\frac{4\pi n e^2}{m_e}\right)^{1/2} \nonumber \\
 & \simeq & (4.7\times 10^8 \ {\rm rad \ s^{-1}}) \ \xi^{1/2} B_{s,15}^{1/2} P^{-1/2} \hat r_2^{-3/2}
 \label{eq:omegap1}
 \end{eqnarray}
\begin{equation}
 \omega_{\rm B} =  \frac{eB}{m_ec} \simeq  (1.8 \times 10^{16} \ {\rm rad \ s^{-1}}) \ B_{s,15} \hat r_2^{-3}. 
 \label{eq:omegaB1}
\end{equation}
It is also informative to evaluate these frequencies at the light cylinder of the magnetar, which has $r=R_{\rm lc} = c/\Omega$ and
\begin{equation}
\hat r_{\rm lc} = (4.8 \times 10^3) P R_6^{-1}.
\label{eq:rlc}
\end{equation} 
This gives
\begin{eqnarray}
 \omega_p^{\rm lc} & \simeq & (1.4\times 10^6 \ {\rm rad \ s^{-1}}) \ \xi^{1/2} B_{s,15}^{1/2} P^{-2} R_6^{3/2}
 \label{eq:omegap2} \\
 \omega_{\rm B}^{\rm lc} & \simeq &  (1.6 \times 10^{11} \ {\rm rad \ s^{-1}}) \ B_{s,15} P^{-3} R_6^{3}. 
 \label{eq:omegaB1}
 \end{eqnarray}
One can see that for the low-frequency EM waves we are interested in, the condition $\omega_0 \ll \omega_p \ll \omega_{\rm B}$ is satisfied throughout the magnetosphere. Therefore the X-mode low frequency waves are transparent to the magnetosphere in all directions for essentially  all frequencies \citep[see also][]{lu19}. 

The above treatment makes the assumption of a cold plasma. Near surface electromagnetic waves need to penetrate through the closed field line region to reach the emission sites in our two scenarios (Fig.\ref{fig:cartoon}). Since the closed field line region of a magnetar is likely populated with inactive non-relativistic particles, the cold plasma dispersion relation treatment can serve the purpose for our discussion. We note that the dispersion relation is more complicated when relativistic plasmas are considered \citep[e.g.][]{arons86,lyubarsky98,melrose17}. Nonetheless, the conclusion that X-mode low-frequency electromagnetic waves are transparent in a magnetar magnetosphere remains valid for such more complicated situations.

\subsection{Cross section and ICS power}\label{sec:ICS}

In the following, we perform a vacuum treatment of the ICS process for simplicity. A more rigorous treatment should consider various plasma effects \citep[e.g.][]{melrose78,rowe95,lyubarsky96,fung04,lyutikov21}. However, a vacuum treatment can catch the essential features of this family of models, as discussed below.

The Compton scattering (in the rest frame of electron) cross section in a strong magnetic field is significantly modified from the Thomson cross section $\sigma_{\rm T}$. The cross section for the two modes of photons read \citep{herold79,xia85}
\begin{eqnarray}
 \sigma' (1) & =& \sigma_{\rm T} \left\{ \sin^2\theta'_i +\frac{1}{2} \cos^2\theta'_i \left [ \frac{{\omega'}^2}{(\omega'+\omega_{\rm B})^2} +  \frac{{\omega'}^2}{(\omega'-\omega_{\rm B})^2} \right ] \right\}, \\
 \sigma' (2) &=& \frac{\sigma_{\rm T} }{2}  \left [ \frac{{\omega'}^2}{(\omega'+\omega_{\rm B})^2} +  \frac{{\omega'}^2}{(\omega'-\omega_{\rm B})^2} \right ].
 \label{eq:sigma}
\end{eqnarray}
where 1 denotes the mode that the electric vector ${\bf E}'$ of the incident wave is parallel to the $({\bf k}, {\bf B})$ plane, and 2 denotes the mode that ${\bf E}'$ is perpendicular to the plane. Here the primed symbols denote in the rest frame of the electron (since electrons move along magnetic field lines, $\omega_{\rm B}$ does not change when the frame changes). In this frame, the photon incident angle $\theta'_i$ is connected to the lab-frame incident angle $\theta_i$ through
\begin{eqnarray}
 \sin \theta'_i &=& \frac{\sin\theta_i}{\gamma(1-\beta\cos\theta_i)}, \\
 \cos\theta'_i &=&  \frac{\cos\theta_i-\beta}{1-\beta\cos\theta_i},
\end{eqnarray} 
where $\gamma$ and $\beta$ are the Lorentz factor and dimensionless speed of the electron, respectively. In the lab frame, the ICS cross section is related to the the rest-frame Compton scattering cross section through $\sigma = (1-\beta\cos\theta_i) \sigma'$ \citep{pacholczyk70}. This finally gives
\begin{eqnarray}
\sigma(1) & = &  \sigma_{\rm T} \left\{ \frac{\sin^2\theta_i}{\gamma^2 (1-\beta\cos\theta_i)} +\frac{(\cos\theta_i-\beta_i)^2}{2(1-\beta \cos\theta_i)} \right. \nonumber \\
&\times & \left. \left [ \frac{{\omega'}^2}{(\omega'+\omega_{\rm B})^2} +  \frac{{\omega'}^2}{(\omega'-\omega_{\rm B})^2} \right ] \right\}, \\
\sigma(2) & = &  \frac{\sigma_{\rm T} }{2}  \left [ \frac{{\omega'}^2}{(\omega'+\omega_{\rm B})^2} +  \frac{{\omega'}^2}{(\omega'-\omega_{\rm B})^2} \right ].
\label{eq:sigma}
\end{eqnarray}

At this point, it is informative to compare the order of magnitude of the two terms in the curly brackets. In the lab frame, $\theta_i$ can be an arbitrary angle, so that all the factors involving $\theta_i$ could be of the order of unity. For our nominal parameters $\gamma \sim 10^{2.5}$ and $\omega_0 \sim (2\pi) 10^4$ and noticing $\omega' \simeq \gamma \omega_0 (1-\beta\cos\theta_i)$, the first term in $\sigma(1)$ is therefore $\propto \gamma^{-2} \sim 10^{-5}$  and the second term (also $\sigma(2)$) is $\propto (\omega' / \omega_{\rm B})^2 \sim (10^{6} / 10^{11})^2 \sim 10^{-10}$ even at the light cylinder where $\omega_{\rm B}$ is the lowest in the magnetosphere.  One can therefore ignore $\sigma(2)$ and the second term of $\sigma(1)$, so that only  \citep[see also][]{qiao98}
\begin{equation}
\sigma(1) \simeq  \sigma_{\rm T} \frac{\sin^2\theta_i}{\gamma^2 (1-\beta\cos\theta_i)} \simeq (6.65\times 10^{-30} \ {\rm cm^2}) \gamma_{2.5}^{-2} f(\theta_i)
\label{eq:sigma2}
\end{equation}
is relevant, where $f(\theta_i) = \sin^2\theta_i/(1-\beta\cos\theta_i)$ is defined. 

The ICS emission power of a single relativistic electron may be estimated as 
\begin{eqnarray}
 P_e^{\rm ICS} & \simeq & \frac{4}{3}\gamma^2 \sigma(1) c U_{\rm ph} \nonumber \\
 & \simeq & (2.1\times 10^{-7} \ {\rm erg \ s^{-1}}) f(\theta_i)  (\delta B_{0,6})^2 \hat r_2^{-2},
 \label{eq:PICS}
\end{eqnarray}
where
\begin{eqnarray}
 U_{\rm ph} & \simeq & \frac{(|{\bf \delta E_0 \times \delta B_0}|)^2}{4\pi} \hat r^{-2} \nonumber \\
 & \simeq & (8.0\times10^6 \ {\rm erg \ cm^{-3}}) (\delta B_{0,6})^2 \hat r_2^{-2}
\end{eqnarray}
is the photon energy density in the emission region, and ${\bf \delta E_0}$ and ${\bf \delta B_0}$ are the electric and magnetic vectors of the low frequency waves near the surface region, which we normalize to a relatively small value $\delta B_0 \sim 10^6 \ {\rm G}$. Comparing with the curvature radiation (CR) emission power of a single electron
\begin{equation}
 P_e^{\rm CR} = \gamma^4 \frac{2e^2 c}{3\rho^2} \simeq (4.6 \times 10^{-15} \ {\rm erg \ s^{-1}}) \gamma_{2.5}^4 \rho_8^{-2}, 
 \label{eq:PCR}
\end{equation}
one can see that the ICS power is much larger, i.e. $P_e^{\rm ICS} \gg P_e^{\rm CR}$, a known fact in pulsar studies \citep{qiao98,zhang99}. This suggests that when coherent ICS operates, the effect of coherent CR is negligibly small.

\subsection{Coherent ICS emission by bunches}

A bunch of leptons with a net charge value of $N_{e,b} e$ can radiate coherently in response to the low frequency waves, with an emitted power of $\sim N_{e,b}^2 P_e^{\rm ICS}$. The observed power is boosted by a factor of $\gamma^2$ because the observer time is shorter by a factor of $(1-\beta\cos\theta) \sim 1/\gamma^2$ (when the angle between the charge motion direction and the line of sight is $\theta < 1/\gamma$) with respect to the emission time. For a total number of bunch number $N_b$, the total luminosity of the emitter may be written as
\begin{equation}
 L \simeq N_b N_{e,b}^2 P_e^{\rm ICS} \gamma^2.
 \label{eq:L}
\end{equation}
One may estimate 
\begin{equation}
N_{e,b} = \zeta n_{\rm GJ} A_b \lambda,
\label{eq:Neb}
\end{equation} 
where $A_b$ is the cross section of the bunch, $\lambda=c/\nu$ is the wavelength, and $\zeta$ is the factor to denote the net charge density with respect to $n_{\rm GJ}$. The most conservative estimate is \citep{kumar17}
\begin{equation}
A_b^{\rm min} = \pi (\gamma \lambda)^2,
\label{eq:Amin}
\end{equation}  
since it describes the causally connected area of a relativistic bunch\footnote{The true $A_b$ can be in principle much larger, up to the area defined by the Fresnel radius i.e. $A_b \simeq \pi x \lambda$, where $x$ is the distance between the emission site and the projected focus of the tangential lines of the emission region field lines.}. Since the ICS mechanism is very efficient, in the following we use this conservative limit. Plugging Eqs. (\ref{eq:ngj}),  (\ref{eq:PICS}), (\ref{eq:Neb}) , and (\ref{eq:Amin}) in Eq. (\ref{eq:L}), one finally gets
\begin{eqnarray}
 L & \simeq & (7.3 \times 10^{38} \ {\rm erg \ s^{-1}}) \zeta^2 N_{b,5} \gamma_{2.5}^6 \nu_9^{-6} \nonumber \\
 &\times& B_{s,15}^2 P^{-2} f(\theta_i) \delta B_{0,6}^2 \hat r_2^{-8}.
 \label{eq:L2}
\end{eqnarray} 
One may compare Eq.(\ref{eq:L2}) with the true FRB luminosity from the observations, i.e. $L^{\rm obs} = L_{\rm iso}^{\rm obs} {\rm max} (\pi/\gamma^2, \pi\theta_j^2)$, where $\theta_j$ is the half opening angle of the FRB jet. For the narrow jet scenario with solid angle defined by the $\gamma^{-1}$ cone, one has $L^{\rm obs} \simeq (3 \times 10^{38} \ {\rm erg \ s^{-1}}) L^{\rm obs}_{\rm iso,43} \gamma_{2.5}^{-2}$. One can see that the coherent ICS mechanism can easily account for the typical FRB luminosity by only requiring a moderate number of $N_b \sim 10^5$ even with the most conservative estimate of the coherent bunch cross section $A_b^{\rm min}$. From Eq.(\ref{eq:L}), one can immediately see that the required $N_b N_{e,b}^2$ for the coherent ICS model is smaller by a factor $P_e^{\rm CR}/P_e^{\rm ICS} \sim 2.2 \times 10^{-8}$ to interpret the same FRB luminosity for the same electron Lorentz factor $\gamma$. 

A bunch with $N_{e,b}$ particles cools more efficiently than single particles since the emission power is $\sim N_{e,b}^2$ times of the individual particle emission power. The bunch quickly loses the total energy, which itself is not large enough to power FRB emission. In order to power FRB emission, a sustained $E_\parallel$ is needed to continuously pump energy to the bunch. Such a scenario is relevant to coherent CR  by bunches \citep{kumar17}. Below we investigate the case for coherent ICS emission. The cooling timescale of the coherent ICS bunches may be estimated as 
\begin{eqnarray}
 t_{c,b} & = & \frac{N_{e,b} \gamma m_e c^2} {N_{e,b}^2 P_e^{\rm ICS}} \simeq (2.1\times 10^{-15} \ {\rm s})  \nonumber \\
 &\times & \zeta^{-1}\nu_9^3 \gamma_{2.5}^{-1} B_{s,15}^{-1} P [f(\theta_i)]^{-1}  \delta B_{0,6}^{-2} \hat r_2^5.
\end{eqnarray}
Since this is much shorter than the typical FRB duration, similar to the bunched coherent CR mechanism, a parallel electric field is also needed to continuously pump energy to the bunch. One may estimate the strength of this $E_\parallel$ by requiring $N_{e,b} e (c t_{c,b}) \simeq N_{e,b} \gamma m_e c^2$, which gives
\begin{eqnarray}
 E_{\parallel} & \simeq & \frac{N_{e,b} P_e^{\rm ICS}}{ec} \simeq (8.6\times 10^9 \ {\rm esu}) \nonumber \\
 &\times &  \zeta \nu_9^{-3} \gamma_{2.5}^{2} B_{s,15} P^{-1} f(\theta_i)  \delta B_{0,6}^{2} \hat r_2^{-5}.
\end{eqnarray}
This is larger than that required for the bunched coherent CR mechanism because of a larger $P_e$. However, in view of the steep dependence on $\hat r$, the required $E_\parallel$ comfortably falls into the range available for magnetars if $\hat r$ is somewhat greater than 100. Several mechanisms may provide this $E_\parallel$ at $\hat r \gg 1$. First, a slot gap near the last open field line region may extend from the neutron surface all the way to light cylinder because the region cannot be filled with electron-positron pairs produced from a pair production cascade -- a consequence of straight line propagation of $\gamma$-rays and curved magnetic field lines \citep[e.g.][]{arons79,muslimov04}. Charge starvation may become more prominent in old, low-twist magnetars \citep{wadiasingh20}. Second, Alfv\'en waves propagating along magnetic field lines would enter a charge starved region at a large enough radius because the charge density may become insufficient to provide the required current (e.g. \citealt{kumar20a,lu20}, cf. \citealt{chen20}). Third, in the current sheet outside of the magnetosphere, an $E_\parallel$ naturally exists in the magnetic reconnection layer  \citep[e.g.][]{lyubarsky20,kala18,philippov18}. 

\section{Salient features of the model}

\subsection{Two physical scenarios}

After discussing some general properties of this model, we discuss two specific scenarios within the framework of the magnetar repeating FRB models. The geometric configurations of the two scenarios are presented in Figure \ref{fig:cartoon}. 

The first scenario is similar to that of \cite{kumar20a} and \cite{lu20}, but with coherent curvature radiation by bunches replaced by the more efficient coherent ICS by bunches. Within this scenario, neutron star crust cracking excites crustal seismic waves, which in turn excite Alfv\'en waves in the magnetosphere. Low frequency electromagnetic waves (effectively fast magnetosonic waves in a high-$\sigma$ medium) are generated and the X-mode waves propagate with a speed close to the speed of light.  The waves are upscattered at a large enough altitude where $E_\parallel$ is developed, either due to charge starvation in the Alfv\'en waves \citep{kumar20a} or a due to the traditional pulsar gap mechanism \citep{arons79,cheng86,muslimov04,wadiasingh20}. Assuming that this emission site is at $\hat r \sim 100$, based on the scalings presented in Section \ref{sec:ICS}, the FRB luminosities are easily interpreted even if the low frequency waves have a relatively low amplitude with $\delta B_{0} \sim 10^6$ G. 

In the second scenario, the emission site is in the reconnection current sheet region just outside of the light cylinder, similar to \cite{lyubarsky20}. However, the radiation mechanism is via coherent ICS by bunches rather than oscillations of colliding magnetic islands within the current sheet as proposed by \cite{lyubarsky20}. Here the bunches could be density fluctuations in the turbulent relativistic particle outflows within the reconnection layer. At this radius, the amplitude of low frequency waves is degraded significantly with respect to that in the near-surface region. Since magnetic reconnection is enhanced by a strong magnetic pulse that is eventually responsible for the FRB, the magnetic energy density is enhanced by a factor of $b \equiv B_{\rm pulse}/B_{\rm wind}$ in the current sheet region \citep{lyubarsky20}. Based on magnetic flux conservation, the cross section of the magnetic tube is also compressed by the same factor $b$. Since the length scale in the ${\bf B}$ direction is not changed, this leads to a local plasma density increased by a factor of $\sim b$. According to Eqs.(1) and (4) of \cite{lyubarsky20}, one estimates $b \sim 4\times 10^4$. As a result, we can estimate the local charge density using Eq.(\ref{eq:n}) with $\hat r = \hat r_{\rm lc}$ (Eq.(\ref{eq:rlc})) and $\zeta \simeq b = 4\times 10^4 b_{4.6}$. Rewriting the equations presented in Section \ref{sec:ICS} with $\hat r = \hat r_{\rm lc}$ (Eq.(\ref{eq:rlc})), one obtains:
\begin{eqnarray}
 U_{\rm ph}^{\rm lc} & \simeq & (3.5\times 10^3 \ {\rm erg \ cm^{-3}}) \delta B_{0,6}^2 P^{-2} R_6^2 \\
 P_{e}^{\rm ICS,lc} & \simeq & (9.3\times 10^{-11} \ {\rm erg \ s^{-1}})  \delta B_{0,6}^2 P^{-2} f(\theta_i) R_6^2 \\
 L^{\rm lc} & \simeq & (4.4 \times 10^{38} \ {\rm erg \ s^{-1}}) b_{4.6}^2 N_{b,9} \gamma_{2.5}^6 \nu_9^{-6} \nonumber \\
 & &  \times B_{s,15}^2 P^{-10} f(\theta_i) \delta B_{0,6}^2 R_6^8 \label{eq:L3} \\
 t_{c,b}^{\rm lc} & \simeq & (1.3 \times 10^{-11} \ {\rm s}) b_{4.6}^{-1} \nu_9^3 \gamma_{2.5}^{-1} B_{s,15}^{-1} \nonumber \\
 && \times P^6 [f(\theta_i)]^{-1} \delta B_{0,6}^{-2} R_6^{-5} \\
 E_\parallel^{\rm lc} & \simeq & (1.4 \times 10^6 \ {\rm esu}) b_{4.6} \nu_9^{-3} \gamma_{2.5}^2 B_{s,15} P^{-6} \nonumber \\
 && f(\theta_i) \delta B_{0,6}^2 R_6^5.
\end{eqnarray}
One can see that the FRB luminosities can be still explained with $\delta B_{0} \sim 10^6$ G but a somewhat larger bunch number $N_b \sim 10^9$. 

Note that in the magnetic reconnection site, the reconnection-driven electric field is perpendicular to ${\bf B}$. This is different from the first scenario where $E_\parallel$ is along the ${\bf B}$ field direction. However, since we do not know the geometric configuration of the FRB central engine, the same FRB may be interpreted by the two scenarios with different viewing geometries: The inner magnetospheric scenario has the viewing angle sweep the open field line region, whereas the reconnection scenario has the viewing angle sweep the equatorial plane (see Fig. \ref{fig:cartoon}). 

\subsection{Energy budget}

In our model, the energy budget to power FRBs is carried by particles rather than by low frequency waves. The latter only carries a small amount of energy. Its role is to provide seed photons for coherent ICS emission to operate. Since $\delta B \propto r^{-1}$ and $B \propto r^{-3}$, with the nominal parameters $\delta B_0 \sim 10^6$ G and $B_s \sim 10^{15}$ G, one estimates $\delta B^{\rm lc} / B^{\rm lc} = 4.8 \times 10^{-6} P R_6^{-1} \ll 1$ at the light cylinder.  As a result, the enhancement of the magnetospheric Thomson cross section (which is relevant when $\delta B > B$, \citealt{beloborodov21})\footnote{In a weak background $\bf B$ field, the Thomson scattering cross section is enlarged by a factor of $a^2$ for strong waves for $a \gg 1$, where $a$ is the amplitude parameter \citep{yangzhang20}. However, with the presence of a strong background field, this effect is suppressed until the wave vector $\delta B$ becomes stronger than the background $B$. } does not occur for the low-frequency waves so that they can propagate freely within the magnetar magnetosphere\footnote{The constraint on the upscattered waves still applies, especially if the emission has a very high luminosity \citep{beloborodov21}. The problem is alleviated if the emission altitude is high (as envisaged in our scenarios) and when radiation pressure \citep{ioka20,wang21} or ponderomotive pre-acceleration of background plasma \citep{lyutikov21b} by the intense FRB emission field are considered.}. The energy of the emitting particles comes from the main magnetic pulse driven by the violent event itself. The energy that powers the FRB is first carried by the Alfv\'en waves, which advect particles to a large altitude where an $E_\parallel$ is developed in a charge-starved region in the magnetar magnetosphere \citep{kumar20a,muslimov04,wadiasingh20} or in magnetic reconnection sites in the current sheet just outside the magnetosphere \citep{lyubarsky20,kala18,philippov18}. Particles are accelerated in these $E_\parallel$ fields so that their ultimate energy comes from the magnetic pulse itself. 

\subsection{Polarization properties}

Since only the X-mode low frequency waves can propagate in the magnetosphere, the incident photon is linearly polarized and has an electric field vector perpendicular to the local magnetic field direction. The outgoing photon in an ICS process is also linearly polarized. According to \cite{herold79}, only the cross section from mode 1 to mode 1 is related to the $\sin^2\theta_i$ term in Eq.(\ref{eq:sigma}). All the other scattering modes (1 to 2, 2 to 1, 2 to 2) involve the $[{\omega'}^2/(\omega'+\omega_{\rm B})^2 + {\omega'}^2/(\omega'-\omega_{\rm B})^2]$ term, which is negligibly small for our problem. The 1 mode has both an X-mode component and an O-mode component. The latter cannot propagate. As a result, our model predicts linearly polarized X-mode emission with a $\sim$ 100\% degree of linear polarization in the X mode. Circular polarization may develop under special conditions \citep{xu00}. This is in general consistent with the FRB observations \citep[e.g.][]{gajjar18,michilli18,cho20,day20,luo20b}. 

The duration of an FRB is defined by the longer of the emission duration and the timescale for the line-of-sight to sweep the bundle of the magnetic field lines where emitting particles flow out \citep{wang19,wang21}. For a slow rotator and an emission site far from the neutron star surface, the polarization angle would be nearly constant across an individual burst, as observed in some repeating FRBs \citep[e.g.][]{michilli18}. For a rapid rotator and an emission site closer to the neutron star surface, the polarization angle may display diverse swing features, as observed in some other repeating FRBs \citep{luo20b}. Our model can account for both patterns.

\subsection{Narrow spectrum and frequency down-drifting}

The characteristic frequency of coherent ICS emission Eq.(\ref{eq:nu}) depends on $\omega_0$, $\gamma$ and $\theta_i$.  For the coherent mechanism to work, particles need to continuously tap energy from $E_\parallel$ so that they are likely emitting in the radiation-reaction-limited regime. As a result, $\gamma$ may remain roughly constant during the emission phase. At the large emission radius as envisaged in the two scenarios, $\theta_i$ varies little as the line of sight sweeps across different field lines. For a seismic oscillation originating from the neutron star crust, the oscillation frequency $\omega_0$ depends on the resonance frequency of the crust which may have a characteristic value. As a result, the characteristic frequency $\nu$ has a narrow value range, unlike curvature radiation which is an intrinsically broad band mechanism \citep{yangzhang18}\footnote{A relatively narrow spectrum can be obtained if oppositely charged particles are spatially separated \citep{yang20c}.}. This is consistent with the narrow-band emission as observed from repeating FRBs  \citep{spitler16,chime-1st-catalog}. 

The characteristic Lorentz factor of electrons may slightly decrease with radius. If so, the outgoing ICS frequency would decrease with radius and give a radius-to-frequency mapping within the ICS model \citep{qiao98}. Alternatively, as the seismic waves damp, $\omega_0$ would gradually decrease with a decreasing amplitude. This would also result in a decreasing $\nu$ of FRB emission with reduced amplitude. Both factors may offer an explanation to the observed sub-pulse frequency down-drifting pattern (also called the ``sad trombone" effect) \citep{hessels19,chime-2ndrepeater,chime-repeaters}\footnote{Other interpretations include radius-to-frequency mapping for curvature radiation \citep{wang19}, decreasing magnetic field strengths in synchrotron maser shocks \citep{metzger19}, or external free-free absorption \citep{kundu21}.}.

\subsection{Required degree of coherence}

Since the ICS emission power of individual electrons is much higher than that of curvature radiation, the required degree of coherence in the bunch is much lower. In the scalings presented in Section \ref{sec:ICS}, the net charge parameter $\zeta$ is normalized to unity. In other words, a bunch with the Goldreich-Julian net charge density can produce the high brightness temperature of FRB emission. In fact, the model even allows the bunch charge density to be below the Goldreich-Julian density, with the luminosity compensated with a somewhat larger total number of bunches, $N_b$, which is currently normalized to a small value. The only requirement to make a bunch is that there are fluctuations of net charge density in space with respect to the background number density \citep{yangzhang18}. This alleviates traditional criticisms on the bunching mechanisms within the context of curvature radiation regarding the formation and maintenance of bunches \citep{melrose78}. Since the mechanism can apply to magnetospheres with a low plasma density (see also \citealt{wadiasingh20}), another major criticism on the bunching mechanism, i.e. the suppression of coherent emission flux due to the dense plasma effect \citep{gil04,lyubarsky21}, is also alleviated. 

\section{Conclusions and discussion}

We have proposed a new family of FRB emission model invoking coherent ICS emission by bunches within or just outside of the magnetosphere of a flaring magnetar. The key assumption is that the oscillations of the magnetar crust would excite oscillations of charges in the magnetosphere near the neutron star surface, which would excite low-frequency electromagnetic waves with an antenna mechanism. The X-mode of these waves could propagate freely in the magnetosphere. Bunched relativistic particles in the outer magnetosphere or in the current sheet region outside of the magnetosphere could upscatter the low frequency waves coherently to power FRBs observed in the $\sim$ GHz band (Eq.(\ref{eq:nu})). For standard parameters, the ICS power of individual particles (Eq.(\ref{eq:PICS})) is much greater than that of curvature radiation (Eq.(\ref{eq:PCR})). As a result, the typical FRB luminosity can be easily reproduced with a low charge density bunch (plasma density of the order of or even lower than the Goldreich-Julian charge density) and a moderate number of bunches (Eqs.(\ref{eq:L2}) and (\ref{eq:L3})). An $E_\parallel$ is needed to sustain the coherent radiation (similar to the curvature radiation mechanism), but such an $E_\parallel$ is expected in charge starved region in the outer magnetar magnetosphere or in the current sheet outside the magnetosphere. This model can account for several observational properties of repeating FRBs, including nearly 100\% polarization degree of emission, constant or varying polarization angle across each burst, narrow emission spectrum, and frequency downdrifting. The low degree of coherence also alleviates several criticisms to the bunched coherent curvature radiation mechanism, including the formation and maintenance of bunches as well as the plasma suppression effect. 

The vacuum treatment discussed in this paper shares some essential features with a broader family of models that invoke relativistic particles scattering off various plasma waves. For example, \cite{lyubarsky96} presented a detailed treatment of induced scattering of relativistic particles off longitudinal subluminal plasma waves and showed that superluminal transverse electromagnetic waves can be efficiently generated to power pulsar radio emission with the characteristic radius-to-frequency mapping. \cite{lyutikov21} studied the scattering of relativistic particles off Alfv\'en wave wigglers and showed a narrow characteristic frequency and a high efficiency, similar to the features of the vacuum model presented in this paper. All these suggest that ICS of relativistic particles against various forms of low-frequency waves could be in general an attractive mechanism to power magnetospheric FRB emission from magnetars. 

\acknowledgments 
The author acknowledges the referee for critical comments that helped to improve the presentation of the paper,  Yuan-Pei Yang, Pawan Kumar, and Wenbin Lu for excellent comments, Guojun Qiao for discussion and encouragement, and Yuanhong Qu for discussion and the help to draw the figure.

\appendix 
\section{Propagation of low frequency waves in a magnetar magnetosphere}

The low frequency electromagnetic waves conjectured in this paper need to propagate through the closed field line region of the magnetar to reach the two FRB emission regions (Fig. \ref{fig:cartoon}). Since the particles residing in the closed field line region of a magnetar magnetosphere are believed to have non-relativistic motion, the dispersion relation of wave propagation may be treated under the assumption of a cold plasma. For a cold, magnetized medium, the conductivity is a tensor $\sigma_{ij}$ so that ${\bf j} = \sigma_{ij} \cdot {\bf E}$, where ${\bf E}$ is the electric field vector of the wave. Define the dielectric tensor 
\begin{equation}
 \epsilon_{ij} = \delta_{ij} + 4\pi i \sigma_{ij} / \omega,
\end{equation}
where $\omega$ is the angular frequency of a wave. The fourth Maxwell's equation becomes ${\bf n} \times ({\bf n} \times {\bf E}) = - \epsilon_{ij} \cdot {\bf E}$, which defines the dispersion relations for wave propagation. In general, defining $\theta$ as the angle between the wave number vector ${\bf k}$ and the magnetic field vector ${\bf B}$, one can write the Maxwell response tensor as \citep[e.g.][]{boyd03}
    \begin{equation}
         \epsilon_{ij} 
        \equiv \left(
        \begin{array}{ccc}
         S-n^2 \cos^2\theta & -i D & n^2\cos\theta\sin\theta \\
         i D & S-n^2 & 0 \\
         n^2\cos\theta\sin\theta & 0 & P-n^2\sin^2\theta
        \end{array}
        \right)
    \end{equation}
where ${\bf B}$ is defined in the $\hat {\bf z}$ direction. Here 
\begin{eqnarray} 
 S & = & \frac{1}{2}(R+L) = 1 - \frac{\omega_p^2 (\omega^2 + \Omega_i \Omega_e)}{(\omega^2-\Omega_i^2)(\omega^2-\Omega_e^2)}, \\
 D & = & \frac{1}{2}(R-L) = \frac{\omega_p^2 \omega (\Omega_i + \Omega_e)}{(\omega^2-\Omega_i^2)(\omega^2-\Omega_e^2)}, \\ 
 R & = & 1 - \frac{\omega_p^2}{(\omega+\Omega_i)(\omega+\Omega_e)}, \label{eq:R} \\
 L & = & 1 -\frac{\omega_p^2}{(\omega-\Omega_i)(\omega-\Omega_e)}, \label{eq:l} \\
 P & = & 1 - \frac{\omega_p^2}{\omega^2}, \label{eq:P}
\end{eqnarray}
where $\omega_p$ is the plasma frequency, $\Omega_e = q_e B/m_e c = - eB/m_ec = - \omega_{\rm B}$ is the electron gyration frequency (which is the Larmor frequency $\omega_{\rm B}$ with a negative sign), and $\Omega_i = q_i B / m_i c$ is the ion gyration frequency (positive sign).  For an ion with atomic number $Z$ and mass number $A$, one has $\Omega_i = Ze B / A m_p c \ll \omega_{\rm B}$. For hydogen, one has $\Omega_i = e B / m_p c$. For an electron positron plasma, one has $\Omega_i = eB/m_ec = -\Omega_e = \omega_{\rm B}$. The general dispersion relation for cold plasma waves is
\begin{equation}
 A n^4 - B n^2 + C = 0,
 \label{eq:dispersion}
\end{equation}
where
\begin{eqnarray}
 A & = & S \sin^2\theta + P \cos^2\theta, \\
 B & = & RL\sin^2\theta + PS(1+\cos^2\theta) \\
 C & = & PRL.
\end{eqnarray}
The propagation of the waves is prohibited at $k=0$ (cut-offs) or at $k \rightarrow \infty$  (resonances) for certain propagation angles. At principle resonances ($k \rightarrow \infty$ and $\theta_{\rm res} = 0$ or $90^{\rm o}$, where the resonant angle $\theta_{\rm res}$  is defined by $\tan^2\theta_{\rm res} = -P/S$), propagation in all directions is prohibited. 

For the case of electromagnetic waves propagating along a magnetic field line, i.e. ${\bf k} \parallel {\bf B}$, the dispersion relations become $n^2 = R$ and $n^2 = L$. Making $R=0$ and $L=0$, one can define two cut-off frequencies for the R-mode and L-mode, respectively, i.e.
\begin{eqnarray}
 \omega_{\rm R} & \equiv &  \left[\omega_p^2+\frac{(\Omega_i-\Omega_e)^2}{4}
        \right]^{1/2}-\frac{(\Omega_i+\Omega_e)}{2} = \sqrt{\omega_p^2 + \omega_{\rm B}^2}, \\  
 \omega_{\rm L} & \equiv & \left[\omega_p^2+\frac{(\Omega_i-\Omega_e)^2}{4}
        \right]^{1/2}+\frac{(\Omega_i+\Omega_e)}{2} = \sqrt{\omega_p^2 + \omega_{\rm B}^2}.
\end{eqnarray}
Hereafter the second equation in each expression is for a pair plasma, where $\Omega_i = - \Omega_e = \omega_{\rm B}$ is adopted. The principle resonances can be obtained from $R \rightarrow \infty$ and $L \rightarrow \infty$, which gives
\begin{eqnarray}
 \omega_{\rm res,R} & = & - \Omega_e = \omega_{\rm B}, \\
 \omega_{\rm res,L} & = & \Omega_i = \omega_{\rm B}.
\end{eqnarray}
Electromagnetic waves are transparent above the cut-off frequencies or below the resonances. So for ${\bf k} \parallel {\bf B}$, the condition for wave propagation in a cold, magnetized plasma is
\begin{equation}
 \omega > \sqrt{\omega_p^2 + \omega_{\rm B}^2}, ~~~ {\rm or} ~~~ \omega < \omega_{\rm B}.
 \label{eq:condition}
\end{equation}

For the case of electromagnetic waves propagating in the direction perpendicular to the field lines,  i.e. ${\bf k} \perp {\bf B}$, one should consider two modes: the O-mode with ${\bf E} \parallel {\bf B}$ and the X-mode with ${\bf E} \perp {\bf B}$. The O-mode is similar to the case as if no magnetic field exists (the electron moving in response of the ${\bf E}$ field does not feel the Lorentz force from ${\bf B}$). Its cut-off frequency is defined by $P = 0$, i.e. $\omega = \omega_p$. The waves cannot propagate at $\omega < \omega_p$.

The X-mode propagation is more complicated. Taking $\theta = 90^{\rm o}$, the dispersion equation (Eq.(\ref{eq:dispersion})) is simplified to 
\begin{equation}
 n^2 = \frac{RL}{S}.
\end{equation}
The cut-off frequencies are again defined by $R=0$ (i.e. $\omega = \omega_{\rm R}$) and $L=0$ (i.e. $\omega = \omega_{\rm L}$), and the resonance is defined by $S = 0$, with the principle resonances defined with an additional condition $\theta_{\rm res} = \pi/2$. This demands $\omega^4 - \omega^2 (\omega_p^2+\Omega_i^2+\Omega_e^2) - \Omega_i \Omega_e (\omega^2 - \Omega_i \Omega_e) = 0$, which defines two principle resonances 
\begin{eqnarray}
\omega^2 = \left(\frac{\omega_p^2+\Omega_i^2+ \Omega_e^2}{2}\right) \left[ 1 \pm \left(1+\frac{4\Omega_i\Omega_e(\omega_p^2-\Omega_i\Omega_e)}{(\omega_p^2+\Omega_i^2+\Omega_e^2)^2} \right)^{1/2} \right].
\end{eqnarray}
For a pair plasma ($\Omega_i=-\Omega_e=\omega_{\rm B}$), one obtains the two (upper and lower) hybrid resonances as
\begin{eqnarray}
 \omega_{\rm uh}^2 & = & \omega_p^2 + \omega_{\rm B}^2, \\ 
 \omega_{\rm lh}^2 & = & \omega_{\rm B}^2.
\end{eqnarray} 
The cut-off bands for X-mode in the case of ${\bf k} \perp {\bf B}$ is $\omega_{\rm uh} < \omega < \omega_{\rm R}$ and $\omega_{\rm lh} < \omega < \omega_{\rm L}$. As shown above, for a pair plasma, one has $\omega_{\rm R} = \omega_{\rm uh} = \omega_{\rm L} = \sqrt{\omega_p^2+\omega_{\rm B}^2}$ and $\omega_{\rm lh} = \omega_{\rm B}$. One therefore draws the conclusion that the cut-off band for a pair plasma in ${\bf k} \perp {\bf B}$ X mode is identical to Eq.(\ref{eq:condition}) derived for the ${\bf k} \parallel {\bf B}$ case.

Since both $\theta=0$ and $\theta=90^{\rm o}$ cases (the latter for X-mode only) have the identical wave propagation condition, Eq.(\ref{eq:condition}) should apply to the oblique case with an arbitrary $\theta$ value for the X-mode waves. O-mode waves cannot propagate below $\omega_p$.


\end{document}